# Evolution of Titan's high-altitude aerosols under ultraviolet irradiation

**Authors:** Nathalie Carrasco[1,2], Sarah Tigrine[1,3], Lisseth Gavilan[1], Laurent Nahon[3], Murthy S. Gudipati[4]

**Affiliations**

[1]LATMOS/IPSL, UVSQ, Université Paris-Saclay, UPMC Univ. Paris 06, CNRS, Guyancourt, France.

[2]Institut Universitaire de France, France.

[3]SOLEIL, l'Orme des Merisiers, St Aubin, BP48, F-91192 Gif sur Yvette Cedex, France.

[4]Science Division, Jet Propulsion Laboratory, Science Division, California Institute of Technology, 4800 Oak Grove Drive, Pasadena, California 91109, USA.

Correspondence to nathalie.carrasco@latmos.ipsl.fr

**Abstract:**

**The Cassini-Huygens space mission revealed that Titan's thick brownish haze is initiated high in the atmosphere at about 1000 km of altitude, before a slow transportation down to the surface. Close to the surface at altitudes below 130 km, the Huygens probe provided information on the chemical composition of the haze. So far we do not have insights on a possible photochemical evolution of the aerosols composing the haze during their descent.**

**We address here this atmospheric aerosol aging process, simulating in the laboratory how solar vacuum-ultraviolet (VUV) irradiation affects the aerosol optical properties as probed by infrared spectroscopy. An important evolution is found, which could explain the apparent contradiction between the nitrogen-poor infrared spectroscopic signature observed by Cassini below 600 km of altitude in Titan's atmosphere, and a high nitrogen content as measured by the Aerosol Collector and Pyroliser of Huygens probe at the surface of Titan.**

**Introduction:**

Only one natural satellite in the solar system hosts a dense atmosphere, Titan the largest satellite of Saturn. With a pressure of about 1.5 bars at the surface, the atmosphere is mainly composed of molecular nitrogen and methane. The photochemistry of these two molecules produces a multitude of heavy organic molecules leading to solid aerosols, responsible for the brownish haze surrounding Titan. The methane concentration is close to 2 % in the stratosphere[1] (40-320 km) which contains most of the mass of the organic haze. NASA's Cassini space mission enabled to observe the aerosols up to 1000 km in Titan's atmosphere thanks to the UVIS spectrometer [2], revealing that Titan's aerosols are created in the ionosphere, a sub-layer of the thermosphere where positive and negative ions are produced through solar ultraviolet radiations and magnetospheric electrons from Saturn [3,4]. The large negative ions have been identified as embryos of aerosols [5]. These nanoparticles aggregate and sediment[6] leading to the haze observed by the Cassini Imaging Science Subsystem[7] and close to the surface by the Descent Imager and Spectral Radiometer of the Huygens probe [8].

Pre-Cassini modeling studies suspected a possible aging of the aerosols through UV radiation or particle impact[9,10] without knowing at that time that aerosols embryos were initiated at altitudes as high as a thousand of kilometers. Yet in the thermosphere highly energetic photons are present: solar Vacuum-UltraViolet (VUV) flux with wavelengths lower than 200 nm. These photons could affect the chemical composition of the aerosols since their formation and as a result the aerosol albedo, impacting the radiative budget of Titan's atmosphere. Our aim is to address experimentally the key issue of the photochemical aging of the aerosols in the thermosphere where VUV photons still reach, knowing that the aerosols descent in the thermosphere, between 1000 and 600 km, lasts one Titanian day (that is about 11 terrestrial days) [6]. We will focus on the evolution of their infrared signatures as a sensitive probe of bond breaking and chemical rearrangement. Those signatures will be compared to the spectra gathered by the Cassini-VIMS instrument in atmospheric layers below the thermosphere.

**Results:**

The goal of this work is to determine if irradiation at wavelengths representative of the thermosphere may impact the chemical composition of the aerosols. The photochemical aging process is simulated in our laboratory experiments by exposing Titan's aerosol analogs to VUV synchrotron radiation (see Fig. 1).

Several analogues of Titan's aerosols can be considered with different properties and chemical structures[11]. In this founding work, we focussed on one type of sample for consistency. The exploration of the same irradiation process on other possible analogues will be valuable in the future. We have therefore synthesized a set of samples similar in size and composition to investigate the effect of the VUV-radiation processing on the freshly formed aerosols. Twenty film samples with 440 ± 20 nm thicknesses have been prepared during the same experiment on Si substrates following the protocol detailed in the Methods section.

In the thermosphere aerosols are submitted to solar energetic VUV photons. The absorption by the atmospheric components leads to an altitude dependent VUV spectrum. Depending on the altitude where the interaction occurs during their descent, the aerosols will be irradiated with photons of different major wavelengths. We have chosen two irradiation wavelengths to experimentally simulate the evolution of the aerosols in the thermosphere: 95 nm representative of hard photons at about 1000 km and 121.6 nm (Lyman-$\alpha$) a major VUV contribution in the solar spectrum penetrating down to 600 km [12]. We have obtained these wavelengths at the VUV DESIRS beamline of SOLEIL synchrotron [13]. The total VUV-UV solar flux density at Titan is about $10^{14}$ photons cm$^{-2}$ s$^{-1}$ (solar irradiance reference spectra at 1 A.U. [14], attenuated by 0.011 due to the distance from the Sun), whereas the DESIRS beamline provides a photon flux density of $10^{16}$ photons cm$^{-2}$ s$^{-1}$. We have therefore chosen irradiation times of a few hours to simulate the irradiation occurring during one Titan-day.

The evolution of the infrared signatures of the organic films was characterized by ex-situ infrared absorption spectroscopy in the 1200-3500 cm$^{-1}$ wavenumber range. The non-irradiated films have been characterized in earlier publications[15]. Vibration modes of double bonds C=C and C=N are found in the 1500-1600 cm$^{-1}$ range. Triple bonds nitrile -C≡N and isonitrile -N≡C have their vibration modes peaked at about 2200 cm$^{-1}$. N-H amine functions show a large absorption band in the 2700-3700 cm$^{-1}$ range. And stretching modes of C-H alkyl signatures leads to sharp bands between 2700 and 3000 cm$^{-1}$ (-CH$_3$ symmetric and asymmetric and -CH$_2$- asymmetric).
Initial spectra preceding the irradiation experiments were taken for each organic thin film and normalized at about 1550 cm$^{-1}$ where the highest absorption value is observed. They are given in Fig 2 as an envelope capturing the dispersion (twice the standard deviation), and. We focus on the evolution of three main chemical functions: nitrile bonds R-CN (and isonitrile R-NC) at ~2200 cm$^{-1}$ (2100-2300 cm$^{-1}$), C-H bonds at ~2950 cm$^{-1}$ (2800-3050 cm$^{-1}$), and amine N-H bonds at ~3200 cm$^{-1}$ (3050-3500 cm$^{-1}$). We find an absorbance dispersion for these three regions of 5% for nitrile bonds, 15% for C-H bonds, and 10% for N-H bonds.

The effect of VUV irradiation is illustrated in Fig. 2 with the comparison of the absorbance spectra of a single sample before and after 24 hours irradiation at 121.6 nm. We observe a general decrease of the targeted signatures, nitrile functions, C-H bonds and N-H bonds. The time-evolution of this effect is reported in Fig. 3 (upper panel). Given the variability of these three absorption features, the decrease becomes significant for only 2 hours of irradiation for the nitrile functions, while more than 20 hours of irradiation are needed to induce a modification of the C-H and N-H bonds.

The samples irradiated at 95 nm evolve similarly, with significant changes observed at shorter irradiation durations only for the nitrile functional group. The kinetics data of the nitrile decrease are compared for the two wavelengths irradiations (95 nm and 121.6 nm) in Fig.3. As expected with the higher energy dose provided at 95 nm, absorption of the nitrile group decreases faster and more significant with 95 nm irradiation than with 121.6 nm. A loss of 17 ± 5% is obtained at 95 nm after 2 hours of irradiation, which is about twice compared to the one observed at 121.6 nm.

Depending on the wavelength, photons penetrate differently the solid sample. This effect is quantified by the penetration depth, D(λ). At this layer thickness, the initial radiation intensity is attenuated by a factor 1/e. The penetration depth of our organic sample is not precisely known. An approximation can be obtained from the imaginary part k of the VUV refractive index obtained for another similar material [16]. Given the equation $D(\lambda) = \frac{\lambda}{4\pi k}$, penetration depths of 26 nm and 11 nm are thus expected for radiation at Lyman-α and 95 nm respectively. The effective layer thickness affected by the radiation is considered at four times the penetration depth with a residual radiation of 2%. The short penetration depth of VUV photons means that these photons are effective up to 100 nm of the 440 nm thick sample during VUV irradiation. Due to the short VUV penetration depth, the material is composed after radiation of two components: ~20% of the total thickness, which has been irradiated and ~80% which did not receive VUV photons. Both components contribute to the total infrared transmission signal of the material after VUV radiation. The experimental 15-20% IR signature attenuation observed on Figure 3 is similar to the thickness contribution of the irradiated layer, showing that the irradiated layer possesses totally different IR properties from the non-irradiated layer.

The 95 nm wavelength has been chosen as a typical wavelength below 100 nm simulating the hard irradiation occurring at about 1000 km of altitude [17]. Yet at about 1000 km, the aerosols initiate their production and reach nanometer-scale dimensions [6,18]. The effect observed with aerosol analogs irradiated at 95 nm will therefore involve a photochemical aging of the aerosols since the very beginning of their formation process. Below 1000 km, we have shown that the evolution process is still active by lower energy radiation such as Lyman-α. A strong modification of Titan aerosol infrared signature is therefore expected through solar VUV aging all along the aerosols sedimentation in Titan's thermosphere.

If we focus on the 2100-2300 cm$^{-1}$ wavenumber range, an evolution of the shape of the structure is detected in addition to a change in the intensity. A deconvolution into four gaussian components enables to characterize this evolution (Fig. 4). Peak spectral position are considered as free parameters during the gaussian fit procedure in order to take into account possible changes in the chemical arrangement from the irradiation. They are found centered at 2110, 2140, 2180 and 2240 cm$^{-1}$ in both cases, with a limited spectral shift between the initial and the irradiated sample (Table 1). According to previous studies [19,20], these components can be attributed to aromatic isonitrile (Ar-NC) at 2110 cm$^{-1}$, aliphatic isonitrile (R-NC) at 2140 cm$^{-1}$, conjugated nitrile (C=C-CN) at 2180 cm$^{-1}$ and aliphatic nitrile (R-CN) at 2240 cm$^{-1}$. There was a debate on the component at 2180 cm$^{-1}$ with a first attribution by Mutsukura and Akita [21] to isonitrile R-NC structures but, since this wavenumber was found at the upper limit of the isonitrile region, it is more compatible with a conjugated nitrile signature [19,20]. A component for aromatic nitriles is found at about 2230 cm$^{-1}$ for similar materials in the literature [19,21], but is negligible in our case, implying that nitrile termination functions are rarely supported by aromatic rings.

The integrated areas of the 2110, 2140 and 2240 cm-1 components evolve upon irradiation at Lyman-α during 24 hrs (see Table 1). The difference observed in the 2100-2300 cm$^{-1}$ region is based on a significant decrease of the single component centered at 2180 cm$^{-1}$, by about 30%. This component attributed to conjugated nitriles is singularly affected by the VUV irradiation at Lyman-α and 95 nm. This result is to be compared with the spectral evolution observed after soft X-ray irradiation of similar aerosol analogs reported in Gavilan et al. [22]. The same conjugated nitrile component at 2180 cm$^{-1}$ decreased after X-ray exposure, but an aromatization was simultaneously observed with an important increase of the aromatic nitrile component at 2230 cm$^{-1}$. This is an important difference between the respective effects of VUV and X-ray irradiations: aromatization of the conjugated nitriles is much smaller under VUV irradiation. A decrease of the conjugate nitrile contribution with minimal effect on the other nitrile or isonitrile signatures in the material involves a loss of this specific nitrile component through a possible breakage of the C-C bond between the double C=C bond and the nitrile termination function. CN and/or HCN losses are expected through this process. Note that in the context of Titan's atmospheric photochemistry, the contribution of X-rays can be neglected, in terms of flux and absorption cross sections, compared to the photochemistry driven by the VUV photons reaching the upper atmosphere of Titan.

The large 2700-3700 cm$^{-1}$ signature of the N-H stretching modes of primary and secondary amine functions decreases after VUV irradiation. This decay is explained by VUV photodissociation of amines (primary and secondary), initiated by Rydberg excitations. Tertiary amines are less photolysed as they stabilise by fluorescence. Primary and secondary amines are therefore depleted, leading to new N-containing functions such as tertiary amines and/or imine. The production of imine functional groups, C=N, is also consistent with the absorbance increase in the 1500 cm$^{-1}$ region. In this process, hydrogen atoms are lost. The consequence is that N-H amine functions are actually consumed, but with no nitrogen loss. After VUV aging, the samples remain nitrogen-rich, but amine-poor.

The sharper bands of the C-H signatures in the 2800-3050 cm$^{-1}$ region show a change in their distribution after irradiation. To characterise the C-H absorption intensities alone, the underlying amine absorption band is substracted after interpolation by a linear model in the 2800-3200 cm$^{-1}$ wavenumber range (Fig. 5). The C-H feature is composed of three main components centered at 2881 cm$^{-1}$ (CH$_3$ symmetric stretching), 2938 cm$^{-1}$ (CH$_2$ asymmetric stretching) and 2965 cm$^{-1}$ (CH$_3$ asymmetric stretching) [19]. After 24hrs of irradiation the intensities of the three components decrease by about 20-30%. This evolution is in agreement with a H-loss of the organic material through VUV photolysis [23]. The decrease is more pronounced for the CH$_2$ groups than the CH$_3$ ones inferring an increase of the CH$_3$/CH$_2$ ratio. This evolution of the CH$_3$/CH$_2$ ratio after irradiation is relevant to the aerosol signature in Titan atmosphere as measured by the Cassini Visible and Imaging Mapping spectrometer [24] at altitude ~ 200 km. A slight shift of a few cm$^{-1}$ also appears for the three components after irradiation typical of a change in the chemical environment of the C-H bonds. A similar positive shift has been observed in the C-H stretching signatures of series of molecules, either through steric effects increasing hydrogen-hydrogen interactions within the material [25] or by the substitution of the adjacent carbon by a nitrogen atom [26]. The interactions with both an opposing hydrogen atom or an adjacent nitrogen lone electron pair hinder the C-H stretch, which requires more energy to occur. The cross-linking of our N-rich material under VUV irradiation would be consistent with both steric and nitrogen effects. Before and after irradiation we note the absence of C-H stretching modes supported by aromatic rings around 3000 cm$^{-1}$ [20]. The presence of signatures above 3000 cm$^{-1}$ in the VIMS data measured at about 200 km (Fig. 5) is therefore not explained by the VUV processing of the aerosols in Titan's thermosphere. Other processes should be invoked to explain this feature such as an aromatic component already present before VUV processing in Titan's thermosphere.

Finally, in the 1500 cm$^{-1}$ region we find an increase in the absorption of the peak at 1570 cm$^{-1}$, as seen in Fig. 2. In this region absorption is due to -C=C- or –C=N- functional groups. We interpret the increase in this absorption as due to radiation-induced cross-linking of polymers in the film. This is the only spectral region that showed an increase in the absorption, indicative of formation of new bonds.

**Discussion:**

The evolution highlighted in this work may also reconcile the contradictory observations on Titan aerosols made by instruments on Cassini and Huygens at different altitudes in the atmosphere.

At the surface, the organic haze has been chemically probed by the ACP instrument during the descent of Huygens (Aerosol Collector and Pyrolyzer). ACP reveals a refractory nucleus rich in nitrogen [27]. The nucleus has been initiated high in the atmosphere: it originates from the nanometric-scale aerosol embryos produced in the ionosphere. The exact chemical composition of the embryos is not given by any direct Cassini measurements in Titan ionosphere, but consistent data support a nitrogen-rich composition involving amine functions. First the fast ion-driven chemistry occuring in Titan ionosphere builds large nitrogen-rich positive and negative ions [28-30]. Then ammonia is an abundant molecule in Titan ionosphere [31] and a possible important gas-phase contributor to the formation of amine functions in the aerosols [32]. And finally analogs of Titan ionospheric aerosols produced with plasma discharge in $N_2$-$CH_4$ gas mixtures also confirm an important amine signatures [15,19].

This aerosol nitrogen content observed at the surface and strongly suspected in the upper atmosphere seems to disagree with observations in the stratosphere made by infrared spectroscopy. The aerosols as seen by Cassini-VIMS in the 2900-3000 $cm^{-1}$ wavenumber region have mainly a signature of aliphatic C-H bonds [33] and the underlying amine contribution expected for N-rich aerosols in this wavenumber range remains hardly detectable.

The experimental evidence of the VUV effect on analogs of Titan upper atmospheric aerosols enables to resolve this apparent contradiction. Chemical growth processes in the upper atmosphere lead to nitrogen-rich species, as revealed by Cassini INMS [30]. The embryos of aerosols produced in this environment are therefore expected to be N-rich, also in agreement with our laboratory analogs showing various nitrogen-containing functions: imines, nitriles and isonitriles, and amines. This work provides evidence that VUV photochemistry with wavelengths below 150 nm could deplete the sensible primary and secondary amine functions of aerosols in the upper atmosphere. This would be consistent with the VIMS and CIRS measurements in the stratosphere, which show no evidence of these functions. The irradiation effect does preserve nitrogen-bearing functionalities that are more strongly bound in the aerosol skeleton, such as tertiary amines, imines and nitriles (except conjugated nitriles). This chemical transformation of the aerosol composition would thus allow the CIRS and VIMS results to agree with the high nitrogen content reported by the Huygens-ACP instrument close to the surface.

**Acknowledgments:**

We are grateful to the general SOLEIL staff for running the facility and providing beamtime under project number 20120579. J.F. Gil is acknowledged for his technical support and the development and design of the sample holder. N. Carrasco and L. Gavilan thank the European Research Council for funding via the ERC PrimChem project (grant agreement No. 636829). We are grateful to B. Fleury for the preliminary IR spectroscopic measurements, as well as P. Pernot helpful discussions. We thank M. Béchard for his help and commitment during his short visit in the team. S. Tigrine acknowledges the University of Paris-Saclay for her thesis funding. MSG's work at the Jet Propulsion Laboratory, California Institute of Technology was done under a contract with the National Aeronautics and Space Administration and funded through NASA-SSW grant "Photochemical Processes in Titan's Atmosphere". L. Nahon acknowledges the support of the Agence Nationale de la Recherche (ANR-07-BLAN-0293).

**Contribution of the authors:**

NC supervised the study, participated to the irradiation experiments, treated the infrared spectroscopic data and drafted the article. All authors discussed the results and commented on the manuscript. LG characterized the sample thickness by ellipsometric measurements. ST participated to the irradiation experience. LN conceived the irradiation set-up, prepared the beamline for the irradiation conditions, and took part in the irradiation experiment. MG was involved in the analysis of the data, its interpretation, reaction mechanisms, and applications to Titan's atmosphere.

**Figures**

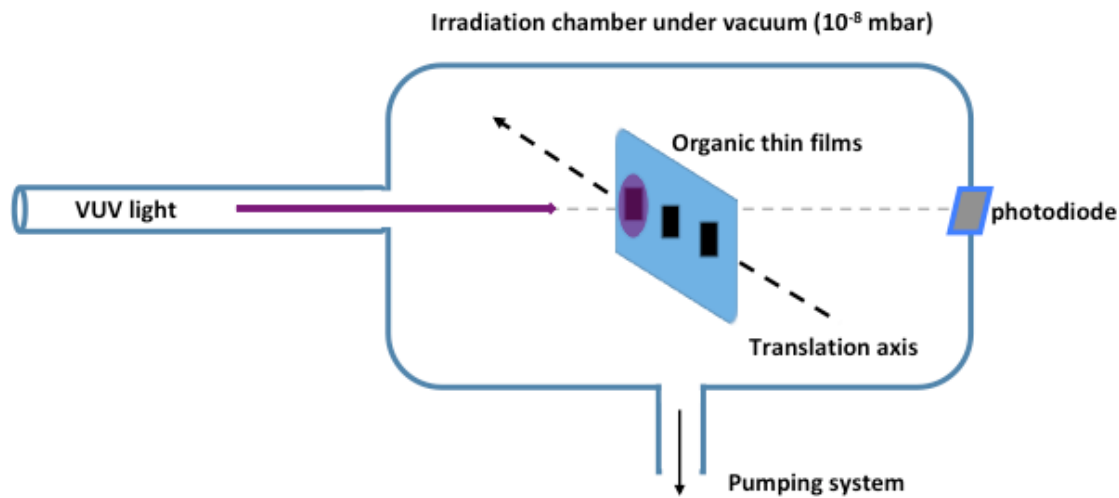

Fig. 1: Sketch of the experiment where the synchrotron VUV beam irradiates Titan's aerosol analogs synthesized as thin organic films of 440 ± 20 nm thickness deposited on Si substrates. Each experiment corresponds to the irradiation of a specific sample at a chosen wavelength and for a given duration. The 11 mm² surface dimension of the films was chosen to ensure a flux density of about $10^{16}$ ph/sec/cm² along with a homogenous irradiation of the sample by the synchrotron beam at the position of the sample holder, without irradiating the adjacent samples.

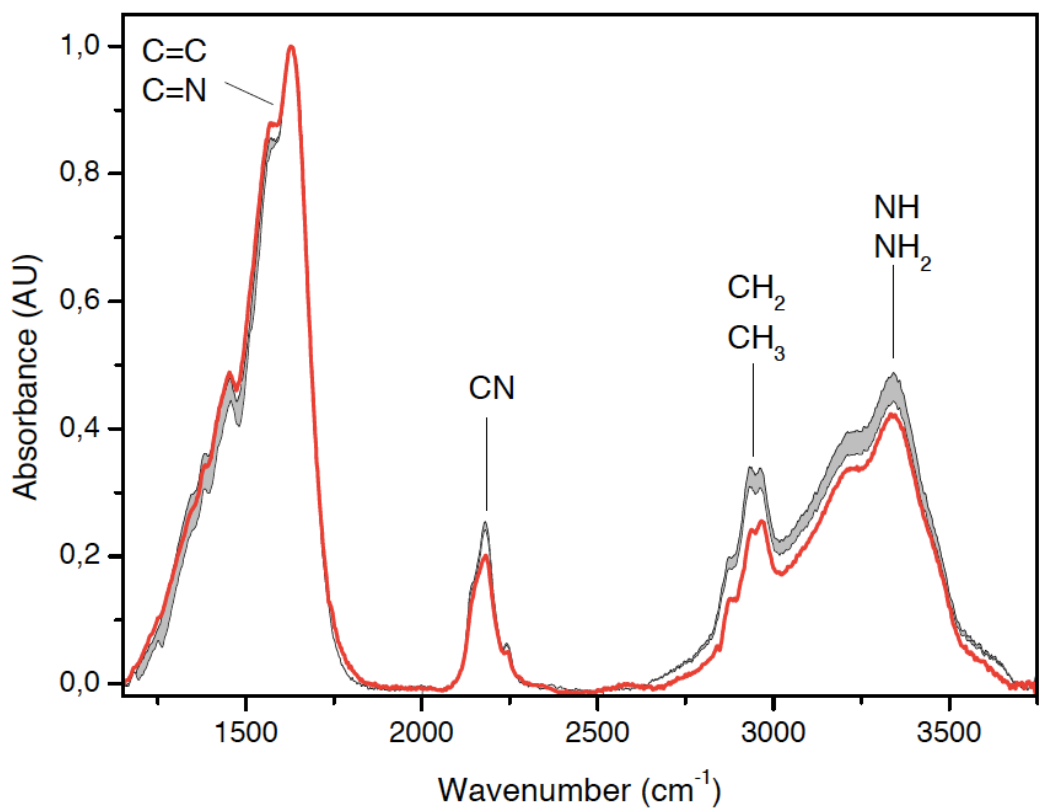

**Fig. 2 : Infrared absorption spectra normalized to the highest absorption value at ~1550 cm$^{-1}$. The grey envelope shows the dispersion (twice the standard deviation) among the twenty non-irradiated samples in the 1200-3500 cm$^{-1}$ range. The red spectrum was recorded after irradiation for 24h by VUV photons at 121.6 nm on a single sample.**

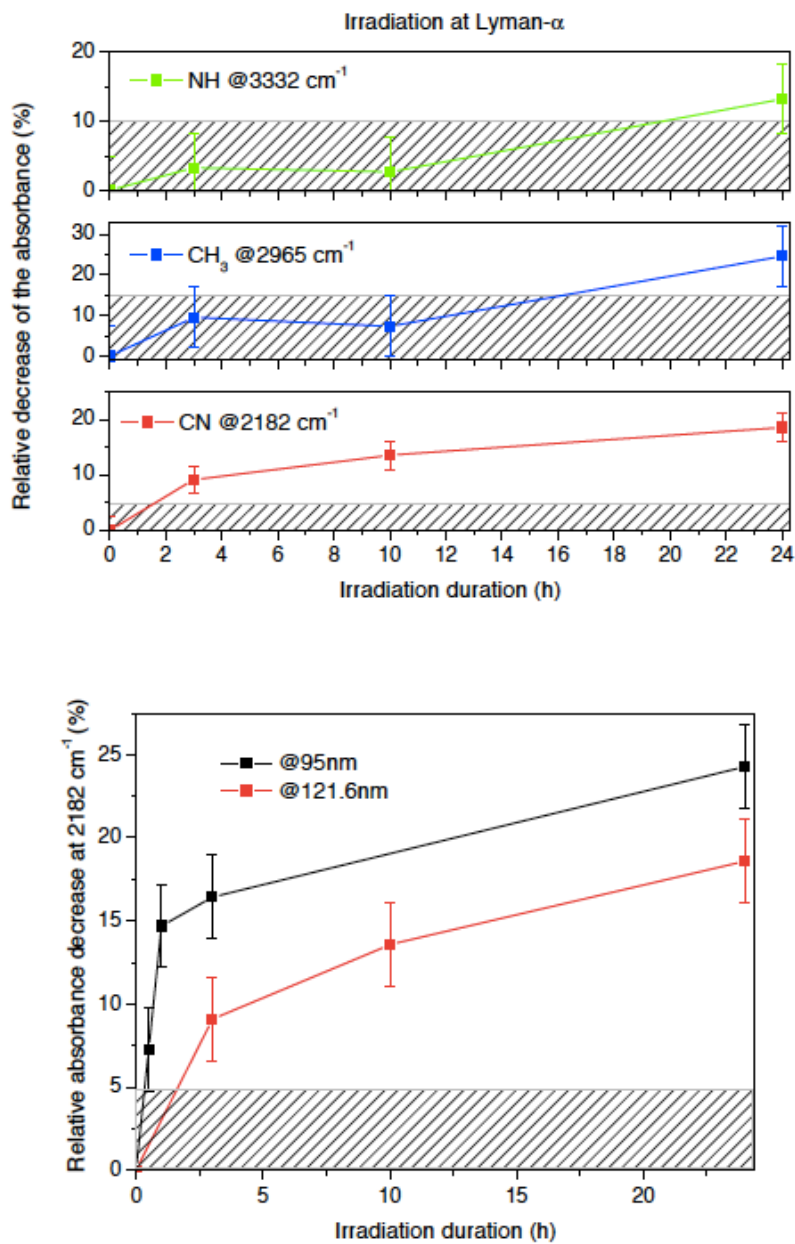

**Fig. 3: (Above)** Irradiation experiments at Lyman-α. Relative decrease with respect to the original intensity of several infrared band maximum intensities *versus* irradiation-time. Hatched boxes indicate the dispersion for non-irradiated samples. Error bars on the data points correspond to twice the standard deviation. Relative decrease becomes significant when larger than the dispersion of the non-irradiated samples.

**(Below)** Decrease of the nitrile intensity at 2182 cm$^{-1}$ under irradiation at 95 and 121.6 nm *versus* irradiation-time. Hatched boxes indicate the dispersion for non-irradiated samples. Error bars on the data points correspond to twice the standard deviation. Relative decrease becomes significant when larger than the dispersion of the non-irradiated samples.

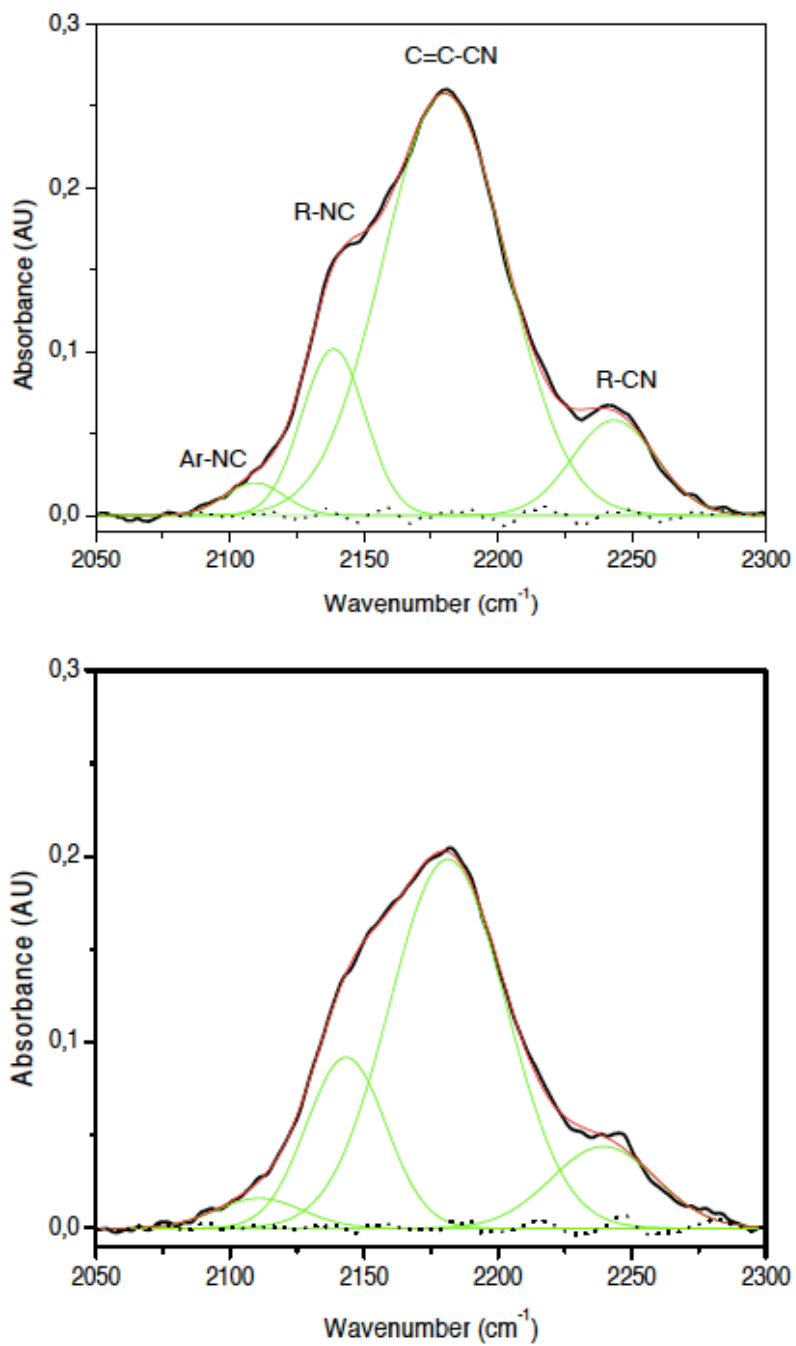

**Fig. 4:** Deconvolution into four Gaussian components of the 2050-2300 cm$^{-1}$ region (above) before irradiation, (below) after 24 hours of irradiation at Lyman-$\alpha$. The black line corresponds to the raw spectrum, the red one to the result from the fitting procedure, the green lines to the four Gaussian components, and in dotted black line the residual.

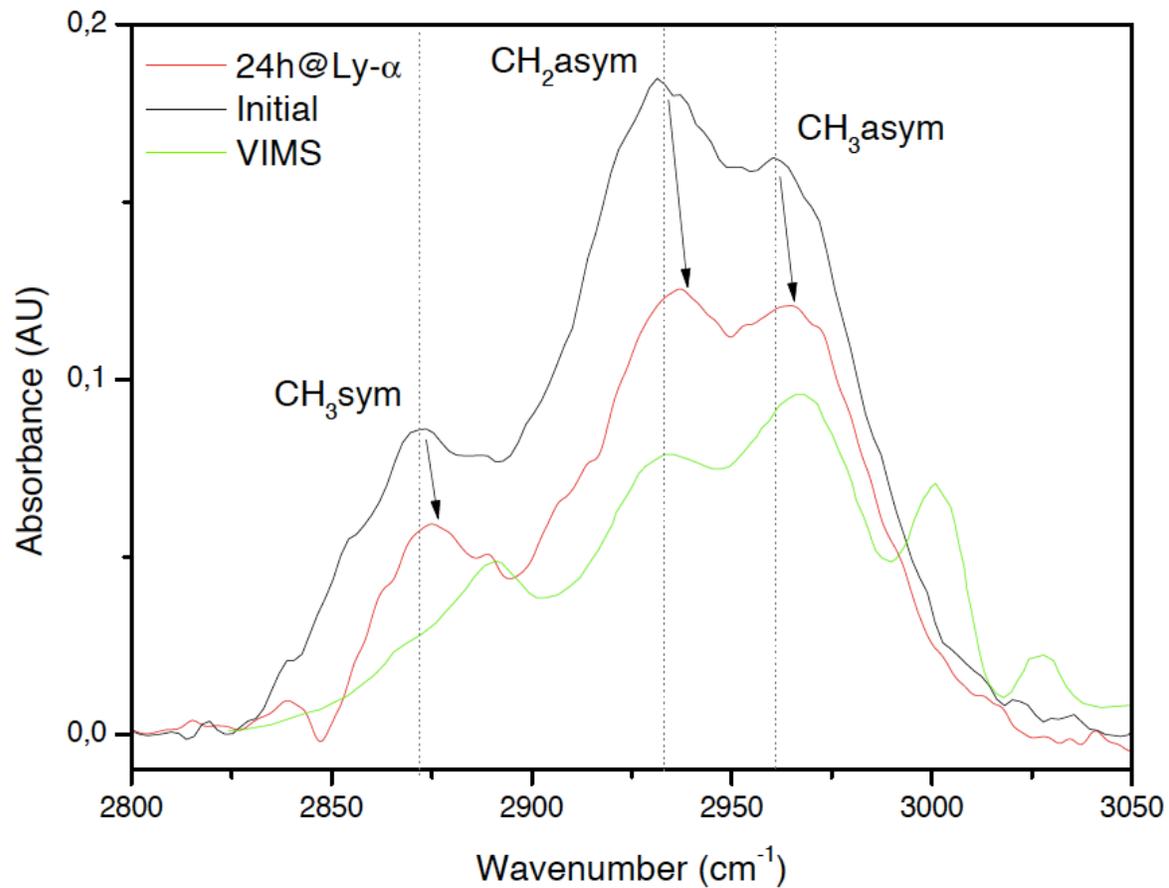

**Fig. 5:** Zoom on the aliphatic carbon band after subtraction of the amine contribution: (Black) before irradiation, (Red) after 24 hrs of irradiation at Lyman-α. (Green) Data from the Cassini Visible and Imaging Mapping spectrometer (VIMS [24]) of Titan's atmosphere at altitude ~ 200 km.

|  | Shift (cm⁻¹) | Integrated areas (arbitrary units) | | |
| --- | --- | --- | --- | --- |
| Gaussian component | | Non-irradiated | Irradiated | Evolution |
| Ar-NC   @2110 cm⁻¹ | -1,7 | 0.55±0.03 | 0.65±0.03 | +0.1±0.1 |
| R-NC    @2140 cm⁻¹ | -4,7 | 3.00±0.15 | 3.42±0.17 | +0.4±0.3 |
| C=C-CN @2180 cm⁻¹ | -1,2 | 14.84±0.74 | 10.84±0.54 | -4.0±1.3 |
| R-CN    @2240 cm⁻¹ | +3,9 | 2.27±0.11 | 2.16±0.11 | -0.1±0.2 |

**Table 1: Effect of the Lyman-alpha irradiation on the integrated areas of the various vibrational structures of the nitrile bands (see deconvolution in Figure 4). Uncertainties on integrated areas of the non-irradiated and irradiated samples correspond to the 5% spectral variability determined in this wavelength range.**

**Methods:**

**Titan's aerosol analogues**
Titan's haze analogs are produced by plasma deposition as thin films onto Silicon windows with a surface of 11 mm² and a thickness of 0.5 mm. In the thermosphere the methane density profile varies between 2 and 10% [1]. Our synthesis has therefore been done by subjecting a 95-5% $N_2$-$CH_4$ gas mixture, representative of Titan's average atmospheric composition in the thermosphere to a radio-frequency electric discharge [2]. The organic film production results from the chemistry induced by electron impact in a $N_2$-$CH_4$ gas mixture using a low pressure of 0.9 mbar radiofrequency capacitively coupled plasma discharge at room temperature [3,4]. The plasma discharge generates electrons that ionize and dissociate $CH_4$ and $N_2$ similarly to the magnetospheric charged particles from Saturn [5] and the VUV photons reaching the atmosphere of Titan [2]. The radicals and ions thus produced react allowing for an efficient macromolecular growth that leads to the production of an organic condensate, which is the laboratory analog of Titan's atmospheric (or thermosphere) aerosols. These analogs produced with a cold-plasma technique are the one simulating best the limited results obtained by the ACP instrument on ESA-Huygens when analyzing the refractory nucleus of the aerosols close to Titan surface [6,7]. This result suggests the importance of reproducing in the laboratory analogs the nitrogen content detected in Titan's aerosols. Previous analysis by UV/VUV spectroscopies [8,9] and solid NMR [10] characterized the chemical structure of our analogs. Those are mainly based on C-N unsaturated bonded units, of which triazine ($C_3N_3$) is an important present component. The samples are considered as analogs of Titan's atmospheric aerosols in terms of replication of the energy source and gas phase composition, but were produced with a higher pressure and temperature than on Titan. The general relevance of laboratory simulations is described in detail in a review paper by Cable et al. [5].

The samples are slightly contaminated by oxygen (~2% of the total elementary composition [11]), in spite of the strict cleaning protocol described in [4] and based on baking under vacuum followed by a pure $N_2$ plasma cleaning. This contamination occurs on the surface of the sample when exposed to ambient air [12]. However this contamination is not an issue here, as the samples infrared spectra treated here are not modified by the small oxygen content, considering the intense signatures of the C-C and C-N bands in the samples. Another source of contamination could be considered during the irradiation process in the vacuum chamber. Radicals and ions released from a sample during its irradiation could interact at the surface of the non-irradiated samples adjacent on the sample holder. No significant thickness evolution is observed after irradiation, showing a negligible quantitative loss or addition of solid material after irradiation. If occurring, a superficial nanometer-scale contamination would not be detected by the infrared transmission spectroscopy diagnosis and is neglected in the present study.

A film thickness of 440 ± 20 nm is found by spectroscopic ellipsometry in the 370–1000 nm spectral range (M-2000V ellipsometer, J.A. Woollam Co) [13]. We verify that the films are similar in size and composition prior the irradiation experiment by comparing their IR signatures. We also make sure that they are homogeneous by recording the IR signatures on different 400 μm ×400 μm spots of the same sample.

**VUV irradiation of the samples**

Considering a solar irradiance reference spectra at 1 A.U. [14], attenuated by 0.011 due to the distance from the Sun, the total VUV-UV solar flux density reaching Titan is about $10^{14}$ photons cm$^{-2}$ s$^{-1}$, and the photon flux density at Lyman-$\alpha$ is $3\times10^9$ photons cm$^{-2}$ s$^{-1}$.

The irradiation duration of the aerosols in the thermosphere corresponds to the fraction of time spent by the aerosols between 1000 and 600 km during which the aerosols are irradiated by the sun. Microphysical models predict that the residence time of the aerosols in Titan thermosphere is about 11 days ($10^6$ s), between their formation at 1000 km and their aggregation in fractal dimensions triggered at 600 km [15]. As a Titan-day lasts about 16 days (1.4 x $10^6$ s), the aerosols are irradiated most of the time during their sedimentation in Titan thermosphere. Considering Titan's rotation and the variation of insolation over Titan's disk, the irradiation time over one day on Titan has to be divided by about a factor of four. During their descent in Titan thermosphere corresponding to about one Titan-day, aerosols receive therefore a total VUV irradiation dose of about $3.5\times10^{19}$ photons cm$^{-2}$.

In our experiments, the 11 mm² sample surface is entirely illuminated by the beam spot of the quasi-monochromatic DESIRS-synchrotron beamline [16]. The samples are positioned vertically on a sample holder at 107 cm from the DESIRS beamline focal point. A high precision manipulator enables to translate from one sample to another. Each sample on the sample holder is irradiated with a specific wavelength (95 or 121.6 nm) and a chosen irradiation-duration. The samples are submitted to the irradiation of the DESIRS monochromator at SOLEIL synchrotron. A gas filter, filled with xenon or argon according to the irradiation wavelength, eliminates the high harmonics of the undulator [17]. In order to maximize the irradiation dose, the grating mirror is settled at the $0^{th}$ order position, where it behaves as a mirror. At a given energy E, the pseudo-Gaussian spectral band of the undulator is transmitted with a spectral resolution $\Delta E/E = 7\%$ at half maximum height. With this configuration, the photon flux density is two orders of magnitude larger than the corresponding UV solar flux density. The photon flux density remains low enough to ensure single-photon interactions with the material. The dose is adjusted by choosing shorter irradiation times of typically a few hours in our experiments. For example during a 24h synchrotron experiment, samples are exposed to a total irradiation dose of $8.6\times10^{20}$ photons cm$^{-2}$, comparable to the total VUV irradiation dose received by the thermospheric aerosols during one Titan-day. The total allocated synchrotron time (4 days) has been distributed in order to measure time-dependent evolutions of the samples under irradiation at both wavelengths. Three samples were irradiated at Lyman-$\alpha$ during 3, 10 and 24h respectively, whereas four samples were irradiated at 95 nm during 30 min, 1, 3 and 24h respectively.

**Infrared absorption spectroscopy**

The evolution of the infrared signatures of the samples was characterized by ex-situ mid-IR absorption spectroscopy in the 1200-3500 cm$^{-1}$ wavenumber range. Initial spectra preceding the irradiation experiments were taken for each organic thin film. All pre- and post-irradiation infrared measurements were performed with the Thermo Scientific Nicolet iN10 MX spectrometer at the SMIS beamline of the synchrotron SOLEIL facility. We used a mercury cadmium telluride (MCT) detector for a transmission analysis and spectral resolution of 4 cm$^{-1}$.

**Data availability**

The data that support the plots within the paper and other findings are available from the corresponding author upon reasonable request.